\documentclass[a4paper,fleqn,usenatbib]{mnras}

\usepackage{newtxtext,newtxmath}
\usepackage[T1]{fontenc}
\usepackage{ae,aecompl}

\usepackage{graphicx, color}
\usepackage{graphicx}	% Including figure files
\usepackage{amsmath}	% Advanced maths commands
\usepackage{amssymb}	% Extra maths symbols

\title[Hyperbolic Orbits in the Solar System]
      {Hyperbolic Orbits in the Solar System: Interstellar Origin or
        Perturbed Oort Cloud Comets?}

\author[A. Higuchi \& E. Kokubo]{
Arika Higuchi,$^{1}$\thanks{E-mail: higuchi.arika@nao.ac.jp}
and Eiichiro Kokubo$^{2}$
\\
$^{1}$RISE Project, National Astronomical Observatory of Japan,
Osawa, Mitaka, Tokyo 181-8588, Japan\\
$^{2}$Division of science,
  National Astronomical Observatory of Japan,
  Osawa, Mitaka, Tokyo 181-8588, Japan}

\date{Accepted XXX. Received YYY; in original form ZZZ}

\pubyear{2019}

\begin{document}
\label{firstpage}
\pagerange{\pageref{firstpage}--\pageref{lastpage}}
\maketitle

\begin{abstract}
  We study the dynamical properties of objects 
  in hyperbolic orbits passing through the
  inner Solar system in the context of two different potential sources:
  interstellar space and the Oort cloud.
  We analytically derive the probability distributions of eccentricity, $e$,
   and perihelion distance, $q$, for each source and estimate the
  numbers of objects produced per 
  unit of time as a function of these quantities.
  By comparing the numbers from the two sources, we assess which
  origin 
  is more likely for a hyperbolic object having a given eccentricity and perihelion distance.
  We find that the likelihood that a given hyperbolic object is of interstellar origin
  increases with decreasing eccentricity and perihelion.  
 Conversely, the likelihood that a hyperbolic object has been scattered from the Oort cloud by a passing star increases with decreasing eccentricity and increasing perihelion.
 By carefully considering their orbital elements, we conclude that both
  1I/2017 U1 'Oumuamua ($e\simeq$ 1.2 and $q\simeq$ 0.26 au) and
  2I/2019 Q4 Borisov ($e\simeq$ 3.3 and $q\simeq$ 2 au)
  are most likely of interstellar origin, not scattered from the Oort cloud.
  However, we also find that Oort cloud objects can be scattered into
  hyperbolic orbits like those of the two known examples, by sub-stellar
  and even sub-Jovian mass perturbers. This highlights the need for better
  characterization of the low mass end of the free-floating brown dwarf
  and planet population.
\end{abstract}


%%%%%%%%%%%%%%%%%%%%%%%%%%%%%%%%%%%%%%%%%%%%%%%%%%

%%%%%%%%%%%%%%%%% BODY OF PAPER %%%%%%%%%%%%%%%%%%
\section{Introduction}
\label{ss:ex_intro}
\begin{figure}
  \begin{center}
    \includegraphics[width=\columnwidth]{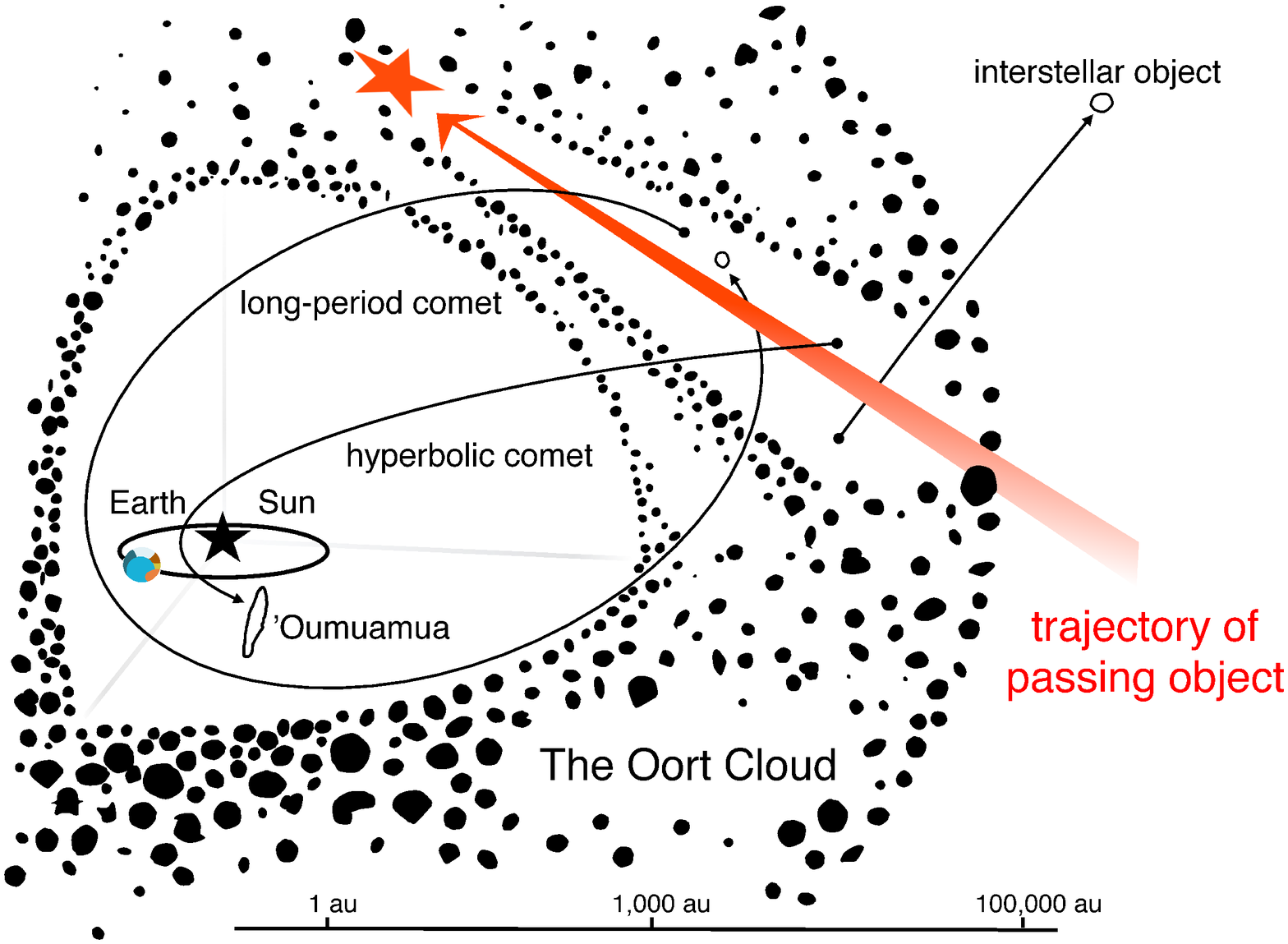}
    \caption{
      Schematic illustration of the penetration of a star through the Oort cloud.
      The star scatters comets away along the trajectory and generates
      long-period comets, hyperbolic comets, and interstellar objects.
    }
    \label{fig:tunnel}
  \end{center}
\end{figure}

The standard formation scenario of planetary systems naturally
suggests that interstellar
space is filled with many planetesimals because exo-giant planets
eject planetesimals
during planet formation, as the planets in the Solar
system did \citep[e.g.,][]{bb:d+04}.
Planetesimals that are almost but not completely ejected
from the planetary
system survive as Oort cloud comets in the planetary system.
Oort cloud comets become observable from Earth
when their perihelion distances become small due to external forces.
For example, when a star penetrates the Oort cloud,
the star drills a narrow tunnel through the Oort cloud
by ejecting the comets within some distance from the star
as described in Figure \ref{fig:tunnel}.
Some of the ejected comets make a last perihelion passage
as their farewell to the Solar system before
becoming fully interstellar objects.
In other words, both  interstellar
space and the Oort cloud are possible as sources of 
objects moving along hyperbolic orbits.

1I/2017 U1 'Oumuamua, (hereafter U1) is the first
highly eccentric ($e\simeq$ 1.2) object identified in the solar system,
with an effective velocity at infinity $V\simeq$ 26 km s$^{-1}$
\citep[e.g.,][]{bb:U1}. 
This velocity cannot be explained by planetary perturbations
because U1 did not encounter any of the planets \citep{bb:m+17}.
Many observations of
U1's shape, thermal properties, colours, absence of cometary
   activity, tumbling rotational state, and non-gravitational
   acceleration have been reported
  \citep[e.g.,][]{bb:j+17, bb:m+17, bb:y+17, bb:b+17, bb:k+17, bb:m+18, bb:f+18, bb:b+18} and are summarized in \citet{bb:b+19}.
Peculiar physical 
properties of U1 include its extremely elongated
or oblate \citep{bb:m19} shape
and its lack of cometary activity.
Together, these properties are unlike those found in other small Solar system objects.
However,  physical peculiarities alone are not enough to exclude the possibility that U1 might be
a Solar system body deflected from the Oort cloud.
We examine this possibility here.
A second hyperbolic object, the comet C/2019 Q4
   (2I/Borisov, hereafter Q4), was discovered by
  G. Borisov on August 30, 2019,
  observing from MARGO, Nauchnij, in the Crimean peninsula.
\footnote{MPEC 2019-R106 : COMET C/2019 Q4 (Borisov)
  https://minorplanetcenter.net/mpec/K19/K19RA6.html}
Soon after that the Q4’s interstellar nature was confirmed.
\footnote{MPEC 2019-S72 : 2I/Borisov = C/2019 Q4 (Borisov)
   https://minorplanetcenter.net/mpec/K19/K19S72.html}
Q4 has a very high eccentricity of
$e = 3.3$, a comet-like appearance and spectrum similar to those of D-type asteroids
\citep{bb:l+19, bb:f+19, bb:jl19}.

\begin{figure}
  \begin{center}
    \includegraphics[width=\columnwidth]{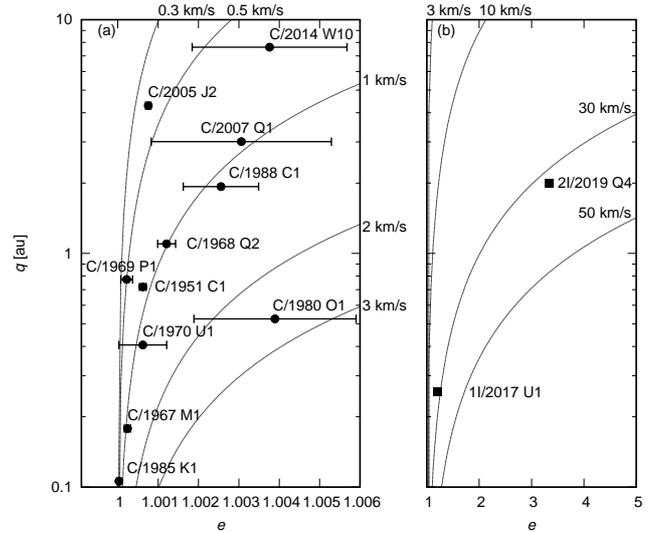}
    \caption{
      Original eccenticities and perihelion distances of
      originally in hyperbolic orbits.
      (a): 11 comets listed in Table. 1 in \citet{bb:kd19}.
      (b): 1I/2017 U1 ('Oumuamua) and 2I/2019 Q4 (Borisov).
      Equi-$V$ curves from eq. (\ref{eq:v}) 
      for $V=$0.3, 0.5, 1, 2, 3, 10, 30, and 50 km s$^{-1}$ are shown with thin dashed curves. }
    \label{fig:kro}
  \end{center}
\end{figure}

While most long-period comets have $e <$ 1, some are known with $e\gtrsim 1$.  
\citet{bb:kd17} calculated the orbits of long-period 
comets carefully taking into account the perturbations from planets and 
the non-gravitational forces to infer their original elements,
defined as the orbital elements at 250 au from the Sun before the perihelion passage
\citep[e.g.,][]{bb:k14, bb:kd17}. 
\citet{bb:kd19} collected data for a full sample of long-period comets discovered 
over the 1801-2017 period and calculated their original orbital elements.
They used the JPL Small Body Database Search Engine
\footnote{https://ssd.jpl.nasa.gov/\_query.cgi}
to construct a complete list of long-period comets
 discovered since 1801, omitting sungrazing comets.
 They found that, in most cases, the comets  followed elliptical (bound)
  orbits prior to their last perihelion.
Figure \ref{fig:kro}(a) shows the original eccentricities $e_{\rm orig}$ and 
perihelion distances $q_{\rm orig}$ of 11 comets in original, marginally hyperbolic orbits
from Table 1 in \citet{bb:kd19}. 
While these comets could also come from the interstellar space,
it is more likely that their eccentricities exceed unity only because of
uncertainties in the astrometry.
In that case, comets in Figure \ref{fig:kro}(a) are dynamically 
the same as other long-period comets, but different from U1 and Q4
shown in \ref{fig:kro}(b).
  
Here, we derive analytically the probability
    distributions of eccentricity, $e$, and perihelion distance,
    $q$, for 
    hyperbolic orbits  derived from either interstellar space or the Oort cloud.
    We estimate the ratio of numbers of objects from the two sources
    and the dependence of this ratio on various parameters
    of the Oort cloud and the intestellar objects. %the local star population.

  In Section 2, we describe the derivation of the likeihood that
  interstellar objects %ISOs
  have a given value of $b$, the impact parameter to the Sun 
  and   $V$, the velocity at infinity.
  Section 3  follows the methodology applied in Section 2
  %the same as Section 2 
  but for the production of comets scattered from the Oort cloud on hyperbolic orbits. %HOCs.
  In Section 4, we plot the probabilities derived in sections 2 and 3
  on the $e$ vs.~$q$ plane and make comparison between
  interstellar objects and hyperbolic Oort cloud comets.
  In Section 5, we compare the expected numbers of
  interstellar objects and hyperbolic Oort cloud comets
  with an assumption that the Solar system recently had an encounter
  with a passing object.
  The properties of a passing object implied by the orbits of U1 and Q4 are discussed in section 6.
  Section 7 gives a summary and discussion.

\section{interstellar objects (ISO)}
Assuming a uniform spatial distribution and
  a Maxwellian velocity distribution,
the number of interstellar objects 
 (hereafter ISOs) encountering the Sun with the velocity at infinity 
 between $V$ and $V+\delta V$ and the impact parameter between $b$ and 
 $b+\delta b$ per time is given by
 
\begin{equation}
  \delta N_{\rm ISO}(V, b)= 2\pi b \delta b  V  \rho_{\rm ISO} p(V)\delta V,
  \label{eq:ex_0}
\end{equation}

\noindent where
$\rho_{\rm ISO}$ is the total number density of ISOs and
$p(V)$ is
a Maxwellian distribution,

\begin{equation}
  p(V)= 
  \sqrt{\frac{2}{\pi}}V^2
  \exp\left(-\frac{V^2}{2a^2}\right)a^{-3},
  \label{eq:ex_pv}
\end{equation}

\noindent where $a=\sqrt{\pi/8}\langle V\rangle$ and $\langle V\rangle$ is the mean velocity.
We assume that ISOs are  planetesimals ejected from planetary
systems by scattering from giant planet(s).
Other fragments might be generated by tidal disruption of planets \citep{bb:c18, bb:r18}
but their expected contribution  is small and neglected here.
We estimate the number density of ISOs generated by stars of spectral type ``i'' as

\begin{equation}
  \rho_{\rm ISO}^i=\rho_{\rm star}^ip_{\rm gp}^in_{\rm OC}^ik_{\rm ISO},
  \label{eq:ex_riso}
\end{equation}

\noindent where
$\rho_{\rm star}^i$ is the number density of the stars,
$p_{\rm gp}^i$ is the probability that the stars have one of more giant planets,
$n_{\rm OC}^i$ is the number of comets in the Oort cloud around each star,
and 
$k_{\rm ISO}$ is set so that $n_{\rm OC}^ik_{\rm ISO}$ gives
the number of ISOs generated by a star of type ``i''.
We use $\rho_{\rm star}$ in \citet{bb:g+01}
and for simplicity set $p_{\rm gp}^i=$0.015, 0.1, and 0
for MK, GFA, and other type stars, respectively
\citep{bb:m+09}.
Assuming that the number of Oort cloud comets
is proportional to the mass of the parent star $m_*^i$, 
we set $n_{\rm OC}^i=n_{\rm OC}^{\rm SS}(m_*^i/m_\odot)$,
where $n_{\rm OC}^{\rm SS}$ is the number of Oort cloud comets
in the Solar system.
We use $m_*^i$ summarized by \citet{bb:r+08},
substituting all the values assumed above and summing over all the 
stellar types, to obtain the total number density of ISOs as

\begin{equation}
  \rho_{\rm ISO}=n_{\rm OC}^{\rm SS}k_{\rm ISO}
  \sum_{i=0}^{13} \rho_{\rm star}^ip_{\rm gp}^i\left(\frac{m_*^i}{m_\odot}\right)
  \simeq \Gamma n_{\rm OC}^{\rm SS}k_{\rm ISO} \;\;,
  \label{eq:ex_riso2}
\end{equation}

\noindent where $\Gamma\simeq 10^{-3}$ [pc$^{-3}$].
Table ~\ref{tb:sp} lists our adopted values.
We assume that the velocity distribution of ISOs is similar to that of 
their parent stars, which is $\langle V\rangle\simeq 50$km s$^{-1}$
as summarized in Table ~\ref{tb:sp} \citep{bb:r+08}.
Substituting Equations (\ref{eq:ex_pv}) and (\ref{eq:ex_riso2}) 
into Equation (\ref{eq:ex_0}), we obtain 
\begin{equation}
  \delta N_{\rm ISO}(V, b) = C_{\rm ISO}\delta V\delta b,
  \label{eq:ex_niso}
\end{equation}
where $C_{\rm ISO}$ is the number density of ISOs with a given $V$ and $b$, written as

\begin{equation}
  C_{\rm ISO}=
  2\pi \Gamma n_{\rm OC}^{\rm SS}k_{\rm ISO}\;V\;p(V)\;b.
  \label{eq:ex_ciso}
\end{equation}

\begin{table}
  \begin{center}
    \begin{tabular}{cccccc}
      \hline\hline
      type& $m_*^i$ [$m_\odot$]&
      $\rho_{\rm star}^i$ [$10^{-3}$pc$^{-3}$]&$p_{\rm gp}^i$
      & $V_*$ [km s$^{-1}$]\\
      \hline\hline
      B0  & 9  &0.06 & 0 &24.6\\
      \hline
      A0  & 3.2  & 0.27& 0 &27.5\\
      \hline
      A5  & 2.1  & 0.44& 0.1 &29.3\\
      \hline
      F0  & 1.7  & 1.42& 0.1 &36.5\\
      \hline
      F5  & 1.3  & 0.64& 0.1 &43.6\\
      \hline
      G0  & 1.1  & 1.52&0.1 &49.8\\
      \hline
      G5  & 0.93  & 2.34&0.1 &49.6\\
      \hline
      K0  & 0.78  & 2.68&0.015 &42.6\\
      \hline
      K5  & 0.69  & 5.26&0.015 &54.3\\
      \hline
      M0  & 0.47  & 8.72&0.015 &50.0\\
      \hline
      M5  & 0.21  & 41.55&0.015 &51.8\\
      \hline
      wd  & 0.9   & 3.0 & 0 &80.2\\
      \hline
      gi  & 4   & 0.43 & 0 &49.7\\
      \hline
    \end{tabular}
    \caption{Stellar parameters used in this paper.
      `wd' and `gi' indicate white dwarfs and giant stars,
      respectively.
      The last column gives the mean heliocentric velocity.
      The values are taken from
      \citet{bb:g+01},
      \citet{bb:m+09}, 
      and \citet{bb:r+08}.
    }
    \label{tb:sp}
  \end{center}
\end{table}

\section{HYPERBOLIC OORT CLOUD COMETS (HOC)}
We first derive the velocity and impact parameter of
a hyperbolic Oort cloud comet (hereafter HOC) against the Sun after an encounter with a 
passing object by using the two-body scattering formula.
Then, we derive the expected number of HOCs for given $V$ and $b$
by taking into account the number density of comets in the Oort cloud.

\subsection{VELOCITY AND IMPACT PARAMETER GIVEN BY A PASSING OBJECT}
We assume that an object that approaches the Sun passes on a straight trajectory.
We describe each encounter of the object with a comet using the
following parameters: $m_*$ and $V_*$, the mass and velocity of the object, 
$b_{\rm Sun}$, the impact parameter of the object to the Sun,
${\bf b}_{\rm HOC}$,
the impact parameter vector from the comet to the object,
and ${\bf r}_*$, the position vector of the object
from the Sun at the moment when the object has the closest approach
to the comet. 
We assume that the comet is not moving relative to the Sun
and $V$ and $b$ of scattered comets are determined only by the perturber.
Also considering $b_{\rm Sun}\gg b_{\rm HOC}$, we approximate the position
vector of the comet from the Sun with ${\bf r}={\bf r}_*$.

The angle between the velocity vectors of the comet to the object
before and after the encounter
$\theta$ is given as a function of
only $V$ and $V_*$.
The angle $\theta$ determines the position of the object at the encounter
so that the comet has an orbit with $b$ after the encounter
(appendix \ref{ss:sa}).
We find $r_*$ that gives $V$ and $b$ as (appendix \ref{ss:fb}),
\begin{equation}
  r_*
  =
  b_{\rm Sun}\left(1-\frac{V^2}{4V_*^2}\right)^{-\frac{1}{2}}.
  \label{eq:ex_rs_kai}
\end{equation}

\subsection{EXPECTED NUMBER OF HOCS PER UNIT OF TIME}
We estimate the number of HOCs encountering the Sun with a velocity
between $V$ and $V+\delta v$ and
an impact parameter between $b$ and $b+\delta b$ per unit of time as

\begin{equation}
  \delta N_{\rm HOC}
  =
  p_{\rm se}
  \delta g  \rho_{\rm OC}(r),
  \label{eq:ex_-3}
\end{equation}

\noindent where 
$p_{\rm se}$ is a probability of having an encounter with an object,
$\delta g$ is an element of volume % a tiny volume for \delta g??
per unit of time (dimensions of $l^3\;t^{-1}$) 
placed at distance $b_{\rm HOC}$ from the passing object,

\begin{equation}
  \delta g
  =
  \frac{8(Gm_*)^2}{V_*b_{\rm Sun}
  } V^{-3}\delta V\delta b,
  \label{eq:ex_g2}
\end{equation}

\noindent and $\rho_{\rm OC}(r)$ is the number density of comets in the Oort cloud at $r$.
The probability $p_{\rm se}$ is 1 if the Solar system just had an encounter with an object,
and if not, $p_{\rm se}=0$.
The  value of $p_{\rm se}$ averaged over the age of the Solar system is discussed 
in section \ref{ss:enc}. 
The element of volume per unit time, $\delta g$, is defined
so that comets contained within $\delta g$
have a velocity between $V$ and $V+\delta V$ and
an impact parameter between $b$ and $b+\delta b$ (appendix \ref{ss:dg}).
We model the distribution of comets in the Oort cloud as
$\rho_{\rm OC}(r)=\bar{\rho}_0 n_{\rm OC}^{\rm SS}r^{-\gamma}$.
Numerical studies show $\gamma \sim 3$ \citep[e.g.,][]{bb:d+04}.
Assuming that the Oort cloud has  inner and outer edges at 
$r_{\rm min}$ and $r_{\rm max}$, respectively, we have

\begin{eqnarray}
  \bar{\rho}_0=
  \left\{ 
  \begin{array}{lll}
    \displaystyle
    \frac{(\gamma-3)}{4\pi}r_{\rm min}^{\gamma-3} &{\rm for} \; \gamma> 3\\
    \left[4\pi \log\left(
      \displaystyle
      \frac{r_{\rm max}}{r_{\rm min}}\right)\right]^{-1} &{\rm for} \; \gamma= 3
  \end{array}
  \right.
  , 
  \label{eq:ex_rhooc}
\end{eqnarray}

\noindent where we assume $r_{\rm min}\ll r_{\rm max}$ for $\gamma>3$.
Substituting Equations (\ref{eq:ex_g2}) and (\ref{eq:ex_rhooc})
into Equation (\ref{eq:ex_-3}), we obtain

\begin{equation}
  \delta N_{\rm HOC}(V,b)=
  C_{\rm HOC}\;\delta V\delta b,
  \label{eq:ex_nocc}
\end{equation}
where $C_{\rm HOC}$ is the number density of HOCs at a given $V$ and $b$ 
and using Equation (\ref{eq:ex_rs_kai})
written as

\begin{equation}
  C_{\rm HOC}=
  \frac{8(Gm_*)^2}{V_*}
  \bar{\rho}_0 n_{\rm OC}^{\rm SS}
  b_{\rm Sun}^{-(\gamma+1)}
  \left(1-\frac{V^2}{4V_*^2}\right)^{\frac{1}{2}\gamma}
  V^{-3}.
  \label{eq:ex_cocc}
\end{equation}

\section{DISTRIBUTIONS OF ECCENTRICITY AND PERIHELION DISTANCE}

We convert the distributions of $V$ and $b$ into those of $e$ and $q$
assuming that all objects move on hyperbolic orbits whose focus is at the Sun (appendix \ref{ss:jacobi}).
The numbers of ISOs and HOCs encountering the Sun with  eccentricity
between $e$ and $e+\delta e$ and the perihelion distance between
$q$ and $q+\delta q$ per time is given by

\begin{equation}
  \delta n(e,q)=CJ \delta e\delta q = \delta n(V, b)J,
    \label{eq:ex_dedq}
\end{equation}

\noindent where 
$\delta n(V,b)$ and $C$ represent $\delta N_{\rm ISO}(V,b)$ or $\delta N_{\rm HOC}(V,b)$ and
$C_{\rm ISO}$ or $C_{\rm HOC}$, respectively,
and $J$ is the determinant of the Jacobian between the $(V,b)$ and $(e,q)$ frames.

\begin{figure*}
  \begin{center}
    \rotatebox{270}{
    \includegraphics[width=0.85\columnwidth]{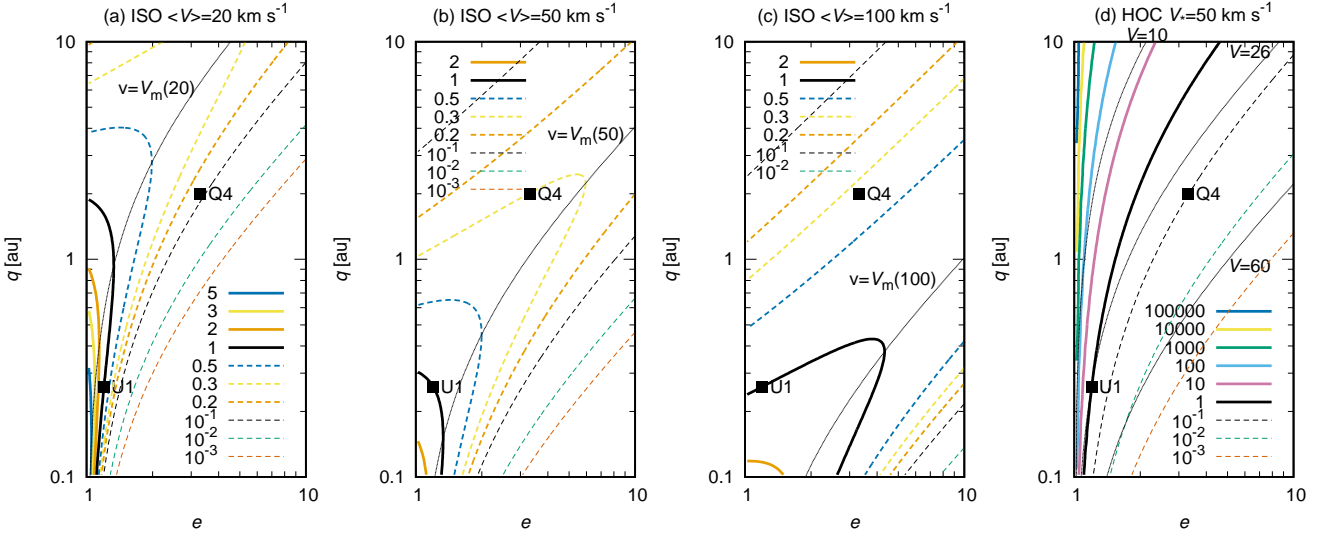}}
    \caption{
      Panels (a), (b), and (c):
      scaled contours of the two-dimensional probability distributions
      plotted on the $e$ vs.~$q$ plane for
      ISOs for $\langle V\rangle=$20 km s$^{-1}$ (a),
      50 km s$^{-1}$ (b), and 100 km s$^{-1}$ (c)
      and
      an equi-$V$ curve for each $V_{\rm m}(\langle V\rangle)$
      (thin black curve).
      Panel (d):
      scaled contours of the two-dimensional probability distributions
      integrated over an encounter with an object with $V_*=$50 km s$^{-1}$
      plotted on the $e$ vs.~$q$ plane and equi-$V$ curves for $V=$10, 26, and 60 km s$^{-1}$
      (thin black curves).
      Black squares in each panel indicate U1 and Q4.}
    \label{fig:2d}
  \end{center}
\end{figure*}

Panels (a), (b), and (c) in Figure ~\ref{fig:2d}
show the contours of the two-dimensional probability distributions
for ISOs obtained from
Equations (\ref{eq:ex_niso}) and (\ref{eq:ex_dedq})
on the $e$ vs.~$q$ plane.
The values of the contours are normalized at U1:
we call this normalized probability $p_{\rm ISO}(e, q)$.
We adopt $\langle V\rangle=20$km s$^{-1}$, 50 km s$^{-1}$, and 100 km s$^{-1}$.

For any value of $\langle V\rangle$ shown in Figure ~\ref{fig:2d},
the probability increases with decreasing $e$ and $q$.
The ridge roughly following the equi-velocity curve 
given by 

\begin{eqnarray}
  V=\sqrt{\frac{Gm_\odot(e-1)}{q}}
  \label{eq:v}
\end{eqnarray}

\noindent for each mode
$V_{\rm m}(\langle V\rangle)=(\sqrt{\pi}/2)\langle V\rangle$ 
is seen, however, the distribution is rather flat.
The probability at the same eccentricity as U1's
but at $q=1$ au, is given by
$p_{\rm ISO}(e=1.2, q=1)\simeq$1.3, 0.34, 0.28, 
for $\langle v\rangle=20$km s$^{-1}$, 50 km s$^{-1}$, and 100 km s$^{-1}$,
respectively.
The probability at Q4's $e$ and $q$
is given by
$p_{\rm ISO}(e=3.3, q=2)\simeq$0.12, 0.30, 0.34
for $\langle V\rangle=20$km s$^{-1}$, 50 km s$^{-1}$, and 100 km s$^{-1}$,
respectively.
This implies that U1's orbit is more typical of ISOs than Q4's.
%This implies that Q4 is less likely to be an ISO compared to U1.
% <- reviewer doesn't like (understand) this sentence.

\begin{figure}
  \begin{center}
    \includegraphics[width=\columnwidth]{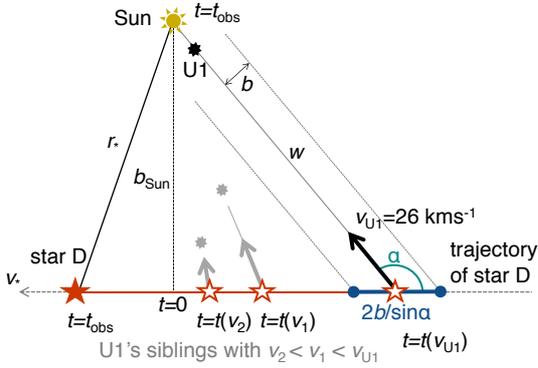}
    \caption{
      Geometry among the Sun, Star D (a passing object),
      U1, and U1's siblings
      plotted on the plane that contains the Sun and the trajectory of Star D,
      as an example of the HOC production.
      Star D passes $b_{\rm Sun}$ at $t=0$.
      U1 arrives at the Sun at $t=t_{\rm obs}$.
    }
    \label{fig:rs}
  \end{center}
\end{figure}

For HOCs, we examine the probability distribution of $e$ and $q$ not per unit of time 
but over an encounter with an object
because it varies with time during the encounter.
We weight Equation (\ref{eq:ex_cocc}) by  $2b/\sin\alpha$,
the path length of the object where it can generate
comets with given $V$ and $b$ (see Figure ~\ref{fig:rs}).
Figure ~\ref{fig:2d}(d)
shows the probability distribution
obtained from Equations (\ref{eq:ex_nocc}) and (\ref{eq:ex_dedq})
on the $e$ vs.~$q$ plane
for HOCs integrated over an encounter with an object 
with $V_*=50$ km s$^{-1}$.
The probability 
diverges at $e=1$ and $q\rightarrow\infty$ ($p_{\rm HOC}\propto q^{2.5}$).
We obtain that $p_{\rm HOC}(e=1.2, q=1)$ is $\simeq$10.
The distribution is steep compared to that of ISOs where $e$ is small.
This result barely changes with $V_*$.
The black dotted lines in Figure ~\ref{fig:2d}(d) show the equi-velocity curves for
$V=10$, 26, and 60 km s$^{-1}$.
Comets on equi-velocity curves arrive at the Sun almost at the same time
  since $b\ll b_{\rm Sun}$.
At $q=1$ au on the $V=26$ km s$^{-1}-$ curve, 
  $p(e=1.76, q=1)\simeq 0.4$.
  This implies that, among the HOCs $V=26$ km s$^{-1}$ that arrive at the Sun
  around the same time,
  U1's $e$ and $q$ are as likely for an origin as HOCs as much as ISOs.
The arrival time
is calculated as $t_{\rm obs}(v)=w/V - (w/\tan\alpha)/V_*\simeq b_{\rm Sun}[1-(3/8)(V/V_*)^2]/V$ 
for $V/V_*<1$, where $w$ is the path length of the HOC (see Figure ~\ref{fig:rs}).
This means that the HOCs with larger $V$ arrive at the Sun earlier
than those with smaller $V$.
%In other words,the outliers of a comet shower are more consistent 
%with U1 than other comets coming after the outliers. %ababababa
In other words, the advance members of a comet shower are more
consistent with U1 than other comets coming after them.
Note that $p_{\rm HOC}(e,q)$ is independent of $m_*$ and $b_{\rm Sun}$.

\section{Ratio of ISO to HOC}
Integration of
Equations (\ref{eq:ex_niso}) and (\ref{eq:ex_nocc})
over ranges of given eccentricity and perihelion distance gives the absolute numbers of
ISOs and HOCs per time.
However, we prefer to discuss  ratio of ISOs to HOCs,  because their absolute numbers
strongly depend on the uncertain size distributions.
In what follows, we have implicitly assumed that ISOs and HOCs have the same
size-frequency distributions, allowing $n_{\rm OC}^{\rm SS}$ to be canceled out.

We define the ratio of the number of HOCs to that of ISOs
for given 
  $e$ and $q$ as
  
\begin{eqnarray}
  \mathcal{H} = \frac{\delta N_{\rm HOC}}{\delta N_{\rm ISO}} 
  = \frac{C_{\rm HOC}}{C_{\rm ISO}},
  \label{eq:h}
\end{eqnarray}

\noindent which tells us which source is more likely   given a particular $e$ and $q$ pair.
We assume that an encounter of the Solar system with an object HOC
occurs and set $p_{\rm se}=1$.

\begin{figure}
  \begin{center}
    \includegraphics[width=0.8\columnwidth]{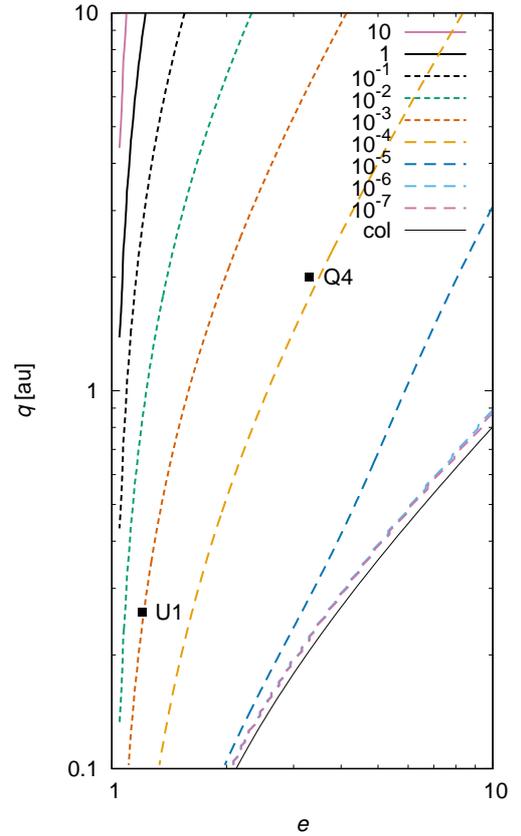}
    \caption{
      Contours of ratios of the number of HOCs to that of ISOs
      on the $e$ vs.~$q$ plane, obtained from Equation (\ref{eq:h})
      for $\gamma=3$, $b_{\rm Sun}=10^4$ au,
      $m_*=10^{-2}\;m_\odot$, $V_*=50$ km s$^{-1}$, and 
      $\langle V\rangle=50$km s$^{-1}$.
      Thin black curve shows eq.(\ref{eq:colm}).
      Black squares indicate U1 and Q4.}
    \label{fig:ratio}
  \end{center}
\end{figure}

Figure ~\ref{fig:ratio} shows contours of $\mathcal{H}$
on the $e$ vs.~$q$ plane 
for $\langle V\rangle=50$ km s$^{-1}$, $b_{\rm Sun}=10^4$ au,
$m_*/m_\odot=10^{-2}$, and $V_*=50$ km s$^{-1}$.
Other parameters are fixed at $\Gamma=10^{-3}$, $k_{\rm ISO}=10$, and $\gamma=3$.
At the $e$ and $q$ of U1 and Q4
in Figure ~\ref{fig:ratio}, $\mathcal{H}\sim 10^{-3}$ and $\sim 10^{-4}$, respectively.
This means that both U1 and Q4 would be less likely to be HOCs,
even if the Solar system had a recent encounter with a passing object as assumed above.
One can easily calculate $\mathcal{H}$ for any
$b_{\rm Sun}$, $m_*$, $\gamma$, and $k_{\rm ISO}$ from Figure ~\ref{fig:ratio} 
as the dependence of $\mathcal{H}$ on
$b_{\rm Sun}$ and $m_*$ is simply $\mathcal{H} \propto b_{\rm Sun}^{-(\gamma+1)}m_*^2
\gamma^{-1}k_{\rm ISO}^{-1}$
(Equation (\ref{eq:ex_cocc})).
For $b_{\rm Sun}=10^3$ au, $\mathcal{H}\sim 10$ and $\sim 1$ 
at the $e$ and $q$ of U1 and Q4, respectively.
The overall trend of $\mathcal{H}$ on the $e$ vs.~$q$ plane
does not change with any of the parameters; diverge at $e=1$ and $q=\infty$.
however, note that there are lower limits of $m_*$
for HOC production defined by
the condition to avoid a collision between a comet and the passing object
(Equation (\ref{eq:colm}))
and the lower limit of $V_*>V/2$ (Equation (\ref{eq:ex_vabs})).
There is no HOC below the curve showing Equation (\ref{eq:colm})
in Figure ~\ref{fig:ratio}.

\begin{figure}
  \begin{center}
    \includegraphics[width=0.8\columnwidth]{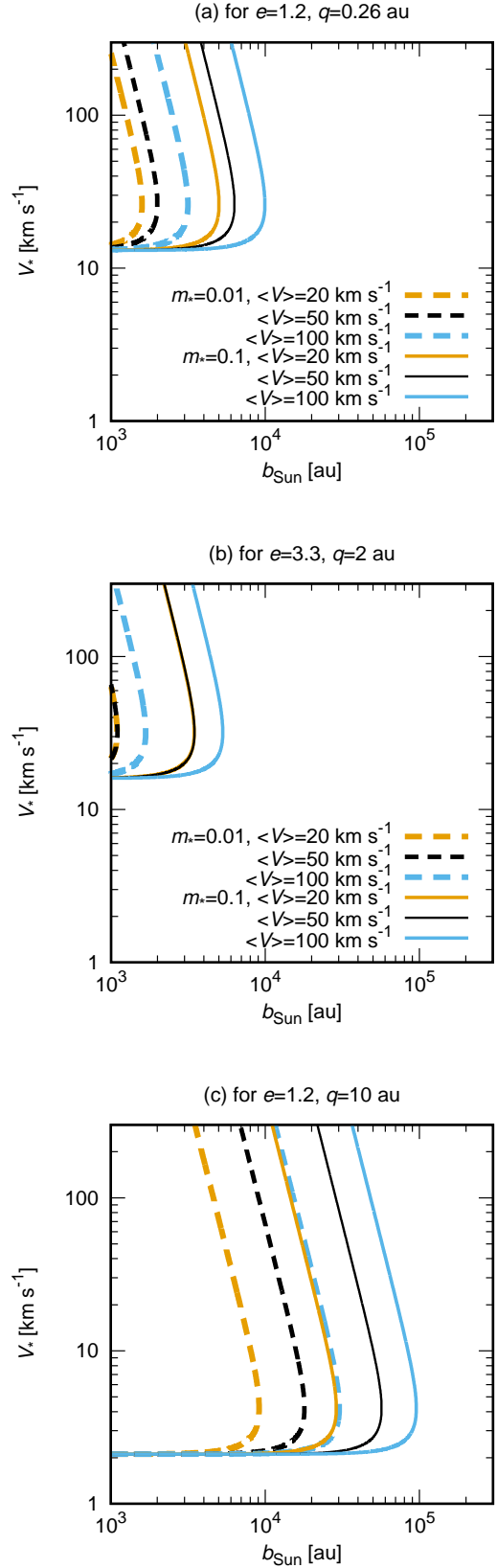}
    \caption{
      Curves for $H_{\rm HOC/ISO}=1$ for given $e$ and $q$
      on the $b_{\rm \odot}-V_*$ plane 
      obtained by solving Equation (\ref{eq:h})$=$1
      for      
      $\langle V\rangle=20$km s$^{-1}$ (orange),
      $50$km s$^{-1}$ (black),
      $100$km s$^{-1}$ (blue),
      and
      $m_*/m_\odot=10^{-2}$ (dashed) and $10^{-1}$ (solid).
      Panels (a), (b), and (c) are for
      $(e,q)$=(1.2, 0.26), (3.3, 2), and (1.2, 10), respectively.
    }
    \label{fig:1}
  \end{center}
\end{figure}

Alternatively to Figure ~\ref{fig:ratio},
    we can derive the condition for a passing object to
    generate hyperbolic minor bodies having an origin in the
    Oort cloud (HOCs) with equal probability to that of being
    interstellar objects (ISOs),
by setting $\mathcal{H}=1$.
Figure ~\ref{fig:1} shows curves for $\mathcal{H}=1$ for given $e$ and $q$
on the $b_{\rm \odot}-v_*$ plane for several $m_*$ and $\langle V\rangle$.
Panels (a), (b), and (c) in Figure ~\ref{fig:1} are for 
$(e, q)=(1.2, 0.26)$, $(e, q)=(3.3, 2)$, and $(e, q)=(1.2, 10)$, 
respectively.
Closed areas 
between the curves and the $y-$ axis in Figure ~\ref{fig:1}
show the range for passing objects to have $\mathcal{H}=1$.
The curve is roughly defined by a horizontal line at the lower limit of $V_*$ 
and a diagonal line for constant $b_{\rm Sun}^{(\gamma+1)}V_*$.
Figures ~\ref{fig:1}(a) and (b) clearly show that
a close encounter with $b_{\rm Sun}\sim 10^3$ au
is required for $\mathcal{H}>1$ if $m_*$ is as small as $\sim$ 0.01 $m_\odot$.
The range for $\mathcal{H}>1$ becomes larger for larger $q$.
If we can observe objects with $q$ up to 10 au, 
an encounter with an object with $m_*$=0.01 $m_\odot$ and $b_{\rm Sun}\sim 10^4$ au 
is enough for $\mathcal{H}>1$ (Figure ~\ref{fig:1}(c)).

\section{Properties of a hypothetical perturber}
\label{ss:enc}

Suppose a hypothetical object, which we will call ``Star D'', 
 scattered an Oort cloud comet onto a hyperbolic object 
 with the velocity at infinity, $V$.
What can we say about the current position and the mass-range of Star D
and about the averaged encounter frequency of the Solar system
 with similar objects?

We assume that Star D is moving along
  a straight trajectory shown in Figure \ref{fig:rs}.
The distance  traveled 
since the instant of time that corresponds to the encounter
  with Star D until now
is estimated from $l=b_{\rm \odot}/\sin\alpha$
and, for Star D, $l_*=lv_*/v$.
Then the distance to the Sun from the current position of
Star D is approximated by Equation (\ref{eq:rs}).
In 3-dimensional space, the geometry of the trajectory of Star D
and ${\bf r}_*$ is axisymmetric about the trajectory of U1.
Therefore, Equation (\ref{eq:rs}) defines a torus-like volume
with a cross section given by the uncertainties of $b_{\rm \odot}$ and $V_*$.
Star D has $r_*\simeq 2 b_{\rm \odot}$ for $V_*=50$ km s$^{-1}$,
where $b_{\rm \odot}\le r_{\rm max}\sim 10^5$ au to penetrate the Oort cloud.

\begin{figure}
  \begin{center}
    \includegraphics[width=0.8\columnwidth]{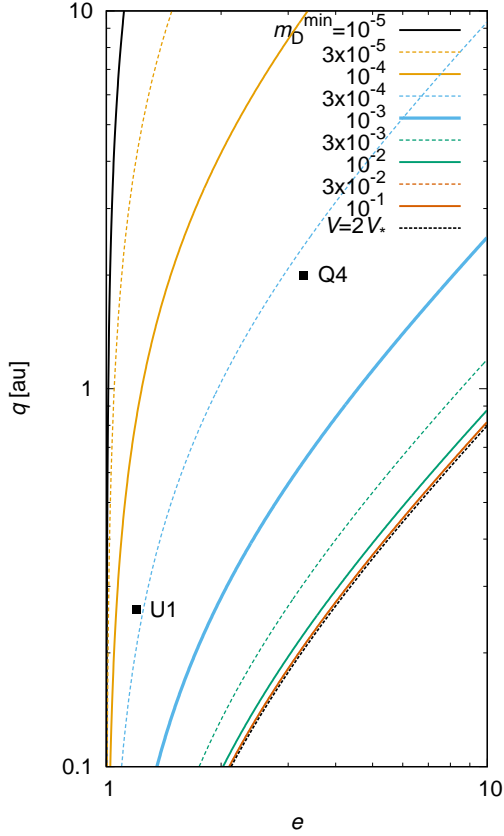}
    \caption{
      Contours of $m_{\rm D}^{\rm min}$
      derived from Equations (\ref{eq:v}) and (\ref{eq:colm}) on the $e$ vs.~$q$ plane
      for $V_*=50$ km s$^{-1}$ and $\rho_*=10^3$ kgm$^{-3}$.
      Black dotted curve shows $V=2V_*$ (No solution below this curve).
      Black squares indicate U1 and Q4.
    }
    \label{fig:mdmin}
  \end{center}
\end{figure}

A lower limit to the mass of Star D, $m_{\rm D}^{\rm min}$, 
is set by the requirement to avoid a collision,
which occurs when impact parameter required to give $V$ 
(Equation (\ref{eq:ex_bocc}))
becomes smaller than the physical radius of Star D.
This leads to

\begin{equation}
  m_{\rm D}^{\rm min}
  =
  \left(\frac{3}{4\pi G^3}\right)^{\frac{1}{2}}
  \rho_*^{-\frac{1}{2}}
  V_*^3
  \left(\frac{4V_*^2}{V^2}-1\right)^{-\frac{3}{4}},
  \label{eq:colm}
\end{equation}

\noindent where $V_*$ and $\rho_*$ are the velocity and density of Star D.
Figure \ref{fig:mdmin} shows the contours of $m_{\rm D}^{\rm min}$
  derived from Equations (\ref{eq:v}) and (\ref{eq:colm}) on the $e$ vs.~$q$ plane
  for $V_*=50$ km s$^{-1}$ and $\rho_*=10^3$ kg m$^{-3}$.
  We have $m_{\rm D}^{\rm min}\simeq 2\times 10^{-4}\;m_\odot$ 
  for the production of both U1 ($V\simeq$26 km s$^{-1}$) 
  and Q4 ($V\simeq$32 km s$^{-1}$).
  This corresponds to $\sim$0.2 Jupiter masses
  For the other ($e \sim$1) comets in Figure \ref{fig:kro}(a), 
  we have $m_{\rm D}^{\rm min}\lesssim10^{-5}\;m_\odot$ (a few Earth masses).
  
An upper limit to the mass of Star D, $m_{\rm D}^{\rm max}$,
can be set by the fact that Star D has not been found by 
the wide-field infrared survey explorer (WISE; \citet{bb:w+10}).
The free-floating planetary-mass object closest
to the Sun is WISE j085510.83-071442.5 \citep{bb:l14}.
Its distance and mass are estimated, respectively, as 2.23 $\pm$ 0.04 pc \citep{bb:le16}
and 3-10 jovian masses, assuming an age of 1-10 Gyr \citep{bb:l14}.
Taking this as a measure of the sensitivity of WISE to nearby sub-stellar objects,
any jovian mass object with the same brightness as the closest one
would have been detected within 1-1.5 pc.
The detection capability of WISE and 
the relation between the brightness and the mass of Star D
are required to give $m_{\rm D}^{\rm max}$.
If $m_{\rm D}^{\rm max}$ is larger than $m^{\rm min}_{\rm D}$,
there is a possibility that U1 is an Oort cloud comet injected by
an object.

The averaged encounter frequency of the Solar system with
the candidates for Star D might be estimated from
that for stars.
Summing up the encounter frequencies of the Solar system with
main-sequence stars, white dwarfs, and giant stars
given in Table 1 in \citet{bb:r+08},
we obtain $\simeq$ 10.5 stellar encounters per Myr within 1 pc.
This is a lower limit because planetary mass objects  have
not been taken into account in \citet{bb:r+08} but may nevertheless scatter comets,
as estimated in Equation (\ref{eq:colm}).
The encounter frequency with such small objects over the age of
the Solar system cannot yet be reliably estimated.
Gravitational microlensing is the only method capable of exploring 
the entire population of free-floating planets down to mars-mass 
objects.
Although this issue is far from well understood \citep{bb:s+11, bb:mr+17},
some authors \citep{bb:mr+17} have given a value for
the frequency of Jupiter-mass free-floating
or wide-orbit planets of 0.25 planets per main-sequence star.
We give $p_{\rm se}=1$ in Equation (\ref{eq:ex_-3})
to compare the numbers of ISOs and HOCs when we have HOCs
(otherwise $\delta N_{\rm HOC}=0$).

\section{SUMMARY AND DISCUSSION}
\label{ss:sd}
We analytically derive the expected distributions
of eccentricity, $e$, and perihelion distance, $q$,
for objects belonging to two distinct populations.
First, we consider initially unbound objects entering
the Solar system from interstellar space (ISOs).
Second, we consider initially bound objects  from
the Oort cloud (HOCs) scattered onto hyperbolic
trajectories   by gravitational interaction with
a passing star (see Figure ~\ref{fig:2d}).
We estimate the numbers of ISOs and HOCs
and evaluate them by using their ratio, $\mathcal{H}$,
on the $e$ vs.~$q$ plane (Figure ~\ref{fig:ratio}).

(1) We find that hyperbolic objects with small $e$ and
small $q$ are the most likely to have an interstellar origin.
Conversely, hyperbolic objects with small $e$ but large $q$
have a higher likelihood of having being scattered
from the Oort cloud.  

(2) Both 1I/'Oumuamua (2017 U1) and 2I/Borisov (2019 Q4)
have orbits most consistent with an interstellar origin.
While an origin by scattering from the Oort cloud cannot
be rejected, 
%this possibility is very small in the absence
this possibility has a very low probability of occurrence in the absence
of a recent and very close stellar encounter,
for which we have no evidence.   

(3) We find that passing bodies of sub-stellar mass
(down to $\sim$0.2 $M_{J}$) are capable of deflecting
Oort cloud comets into hyperbolic orbits like those of
1I/'Oumuamua (2017 U1) and 2I/Borisov (2019 Q4).  \\

Future observations of two kinds are needed to provide an improved
understanding of the dynamics and origin of hyperbolic objects
in the Solar system.
First, the distribution of orbital elements of such bodies,
especially in the eccentricity vs.~perihelion distance plane,
will help determine the ratio of interstellar to scattered
Oort cloud sources.
Second,  measurements of the abundance and distribution of
sub-stellar (even sub-Jupiter) mass perturbers near
the Sun are needed to  quantify the role of scattering
from the Oort cloud.

\section*{ACKNOWLEDGEMENTS}
We thank the anonymous referee for helpful comments on the paper,
Dimitri Veras for carefully reading the first version of
the manuscript,
and David Jewitt for his comments that greatly improved
the quality of this paper.

%%%%%%%%%%%%%%%%%%%%%%%%%%%%%%%%%%%%%%%%%%%%%%%%%%

%%%%%%%%%%%%%%%%%%%% references %%%%%%%%%%%%%%%%%%

% the best way to enter references is to use bibtex:

%\bibliographystyle{mnras}

% alternatively you could enter them by hand, like this:
% this method is tedious and prone to error if you have lots of references
%\begin{thebibliography}{99}
%\bibitem[\protect\citeauthoryear{author}{2012}]{author2012}
%author a.~n., 2013, journal of improbable astronomy, 1, 1
%\bibitem[\protect\citeauthoryear{others}{2013}]{others2013}
%others s., 2012, journal of interesting stuff, 17, 198
%\end{thebibliography}

%%%%%%%%%%%%%%%%%%%%%%%%%%%%%%%%%%%%%%%%%%%%%%%%%%

%%%%%%%%%%%%%%%%% APPENDICES %%%%%%%%%%%%%%%%%%%%%
\appendix
\section{Derivation}
\subsection{Scattering Angle}
\label{ss:sa}
We assume a passing object that approaches the Sun on a hyperbolic orbit.
Using non-rotational coordinates centered on the object having a
given hyperbolic orbit defined by $V_{*\infty}$ and $b_{\rm \odot}$, a comet
encounters the object with the velocity and impact parameter $V_*$ and
$b_{\rm HOC}$, respectively.
The velocity of the object at the moment of the closest approach
%encounter
is given by
\begin{equation}
  V_*=\sqrt{V_{*\infty}^2+\frac{2GM}{r_*}},
  \label{eq:ex_vs2}
\end{equation}
where $M=m_\odot+m_*$.
The angle between the velocity vectors of the comet before and after
the encounter $\theta$ is given by
\begin{equation}
  \tan\frac{\theta}{2}=\frac{Gm_*}{V_*^2 b_{\rm HOC}}.
  \label{eq:ex_tan_2}
\end{equation}
Then, the velocity of the comet to the Sun after the encounter is expressed as
\begin{equation}
  V=\sqrt{V_*^2+V_*^2-2V_*V_*\cos\theta}
  \;=\;2V_*\sqrt{\frac{\tan^2\frac{\theta}{2}}{1+\tan^2\frac{\theta}{2}}},
  \label{eq:ex_vabs}
\end{equation}
For $\theta\ll 1$,
\begin{equation}
  V = \frac{2Gm_*}{V_*b_{\rm HOC}},
  \label{eq:ex_vabs_ia}
\end{equation}
which is the velocity change given by the impulse approximation.
Scattering that gives the velocity as large as the U1's, which is $\sim V_*$, cannot be dealt with using the impulse approximation.
Equation (\ref{eq:ex_vabs}) gives
\begin{equation}
  \tan\frac{\theta}{2}=\left(\frac{4V_*^2}{V^2}-1\right)^{-\frac{1}{2}}.
  \label{eq:ex_tan_2_2}
\end{equation}

Next, we choose the non-rotating Cartesian coordinates centered
on the Sun such that the $x$ axis is anti-parallel to ${\bf V}_*$,
the $z$ axis is anti-parallel to the angular momentum vector
of the object,
and the $y$ axis is perpendicular to the $x$ and $z$ axes.
The velocity vector of the comet after the encounter is expressed as
${\bf V}=V(\cos \alpha\cos \beta,\sin \alpha\cos \beta,-\sin \beta)$,
where $\alpha=(\pi+\theta)/2$ is the angle between the $x$ axis and ${\bf V}$
and $\beta$ is the angle between ${\bf b}_{\rm HOC}$ and the reference plane.
Using Equation (\ref{eq:ex_tan_2_2}), we have
\begin{equation}
  \sin\alpha  =\left(1-\frac{V^2}{4V_*^2}\right)^{\frac{1}{2}}.
  \label{eq:ex_sinalpha2}
\end{equation}

\subsection{Object Position}
\label{ss:fb}
For the comet to have a trajectory with $b$ after the encounter,
the position of the object during the encounter must be determined.
Let the angle between ${\bf r}_*$ and the $x$ axis be $\alpha_*$.
From the conservation of angular momentum, 
\begin{equation}
  \sin\alpha_*=\frac{b_{\rm \odot}V_{*\infty}}{r_*V_*}.
  \label{eq:ex_sinalphas}
\end{equation}
By combining Equations (\ref{eq:ex_sinalpha2}) and (\ref{eq:ex_sinalphas})
and using Equation (\ref{eq:ex_vs2}), we find $r_*$ that gives $V$ and $b$ as 
\begin{equation}
  r_*
  =
  b_{\rm \odot}\left(1-\frac{V^2}{4V_{*\infty}^2}\right)^{-\frac{1}{2}}S,
  \label{eq:rs}
\end{equation}
\begin{equation}
  S=
  \left[1
    +\left(\frac{GM}{V_{*\infty}^2b_{\rm \odot}}\right)^2\left(1-\frac{V^2}{4V_{*\infty}^2}\right)^{-1}
    \right]^{\frac{1}{2}}
  -\frac{GM}{V_{*\infty}^2b_{\rm \odot}}\left(1-\frac{V^2}{4V_{*\infty}^2}\right)^{-\frac{1}{2}}\simeq 1,
  \label{eq:ex_rskai_s}
\end{equation}
where $V_{*\infty}\ne 0$.
The assumption of $S=1$ corresponds to the approximation that the
trajectory of the object is not hyperbolic but a straight line.
We give $S=1$ and $V_{*\infty}=V_*$ 
since this is true in almost all cases in this paper.

\subsection{Derivation of $\delta g$}
\label{ss:dg}
The tiny volume $\delta g$ is defined with the following equation so
that comets contained within $\delta g$ have $V$ and $b$;

\begin{equation}
  \delta g
  =
  |2\pi b_{\rm HOC}\delta b_{\rm HOC}\times V_*\times\frac{\delta\beta}{\pi}|,
  \label{eq:ex_g}
\end{equation}

\noindent where the ring-area with the radius of $b_{\rm HOC}$ decides $V$ 
and $\delta\beta$ gives the direction of $V$ to meet the Sun with $b$.
Using $b_{\rm \odot}\gg b_{\rm HOC}$, the relation between $\beta$ and $b$ is
\begin{equation}
  \beta\;\simeq\;\sin\beta=\frac{b}{r_*\sin\alpha}.
  \label{eq:beta}
\end{equation}
Substituting Equation (\ref{eq:ex_rs_kai}) into Equation (\ref{eq:beta})
and carrying out the differentiation, we obtain
\begin{equation}
  \delta\beta =  \frac{\delta b}{r_*\sin\alpha}
  =\frac{\delta b}{b_{\rm \odot}
  }.
  \label{eq:ex_dbeta}
\end{equation}
The explicit expression of $b_{\rm HOC}$ is given from
Equations (\ref{eq:ex_tan_2}) and (\ref{eq:ex_tan_2_2}) as
\begin{equation}
  b_{\rm HOC}=\frac{Gm_*}{V_*^2}\sqrt{\frac{4V_*^2}{V^2}-1}.
  \label{eq:ex_bocc}
\end{equation}
By carrying out the differentiation of Equation (\ref{eq:ex_bocc}), we obtain
\begin{equation}
  \delta b_{\rm HOC}
  =-4\frac{(Gm_*)^2}{V_*^2} V^{-3}b_{\rm HOC}^{-1}\delta V.
  \label{eq:ex_dbocc}
\end{equation}
Substituting Equations (\ref{eq:ex_dbeta}) and (\ref{eq:ex_dbocc}) into Equation (\ref{eq:ex_g}),
we obtain $\delta g$ as a function of $V$ and $b$ (Equation (\ref{eq:ex_g2})).

\subsection{Coordinate Transformation from Impact Parameters to Orbital Elements}
\label{ss:jacobi}
From the relation of $V=\sqrt{Gm_\odot(e-1)/q}$ and $b=q\sqrt{(e+1)/(e-1)}$,
the determinant of the Jacobian between the $(V, b)$ and $(e, q)$ frames
is calculated as
\begin{eqnarray}
  J=\left|
  \begin{array}{cc}
    \frac{\partial V}{\partial e} &  \frac{\partial V}{\partial q}\\
    \frac{\partial b}{\partial e} &  \frac{\partial b}{\partial q}\\
  \end{array}
  \right|
  =
  \frac{1}{2}\sqrt{\frac{Gm_\odot}{q(e+1)}}\frac{e}{e-1}.
  \label{eq:ex_jacobian}
\end{eqnarray}
%%%%%%%%%%%%%%%%%%%%%%%%%%%%%%%%%%%%%%%%%%%%%%%%%%

% Don't change these lines
\bsp% typesetting comment
\label{lastpage}
\end{document}